\documentstyle[11pt,paspconf]{article}

\begin{document}

\title{HST imaging of two $z>4$ radio galaxies}
\author{M.\ Lacy}
\affil{Department of Physics, Oxford University}

\begin{abstract}
We have imaged 8C 1435+635 ($z=4.25$; Lacy et al.\ 1994; Spinrad, Dey 
\& Graham 1995) and 6C 0140+326 ($z=4.41$; Rawlings et al.\ 1996) in both 
continuum and Ly$\alpha$ plus continuum bands with the HST. Our images show
patchy distributions of continuum and line emission with a tendency
for the peaks in the two types of emission to be offset. 
When compared to Keck $K$-band images, it seems likely that the 
presence of dusty neutral gas is strongly influencing the 
UV continuum and Ly$\alpha$ emission. The radio and UV emission are not 
particularly well-correlated, except for a general tendency towards 
radio-optical alignment. It is suggested that at least some of 
the observed radio-optical alignment is produced
as dust is cleared from the radio lobes by shocks associated by the 
radio source. If true, this means that 8C1435+635 could be hosting a luminous, 
galaxy-scale starburst, as suggested by submm observations.
\end{abstract}


\section{Observations}

\subsection*{8C 1435+635}

This object was observed in the F814W and F622W
filters (F622W includes the Ly$\alpha$ line). Using the Ly$\alpha$
flux of Spinrad et al.\ 1995 we have produced a ``pseudo-Ly$\alpha$'' image
by scaling and subtracting the F814W image from the F622W one. 
In Fig.\ 1 we show the F814W image, 
the ``pseudo-Ly$\alpha$'' image and the MERLIN 
5GHz radio map superposed on the Keck $K$-band image of 
van Breugel et al.\ (1998). (The radio-optical registration should be accurate
to $\approx 1$ arcsec.) The Ly$\alpha$ peaks in the southern 
radio lobe, slightly offset from a faint patch of continuum emission. 
The central region has a (resolved) peak of UV emission surrounded by 
diffuse material. The $K$-band image shows that the rest-frame 
blue continuum is diffuse
and covers the whole area enclosed by the radio lobes. The UV emission 
in the centre of the source is offset to the SW, but that in the southern
lobe is closer to the major axis of the $K$-band emission. There is a 
suggestion of a bar of $K$-band 
emission perpendicular to the radio axis across the centre of the source. 

\begin{figure}
\begin{picture}(400,500)
\put(-200,-230){\includegraphics{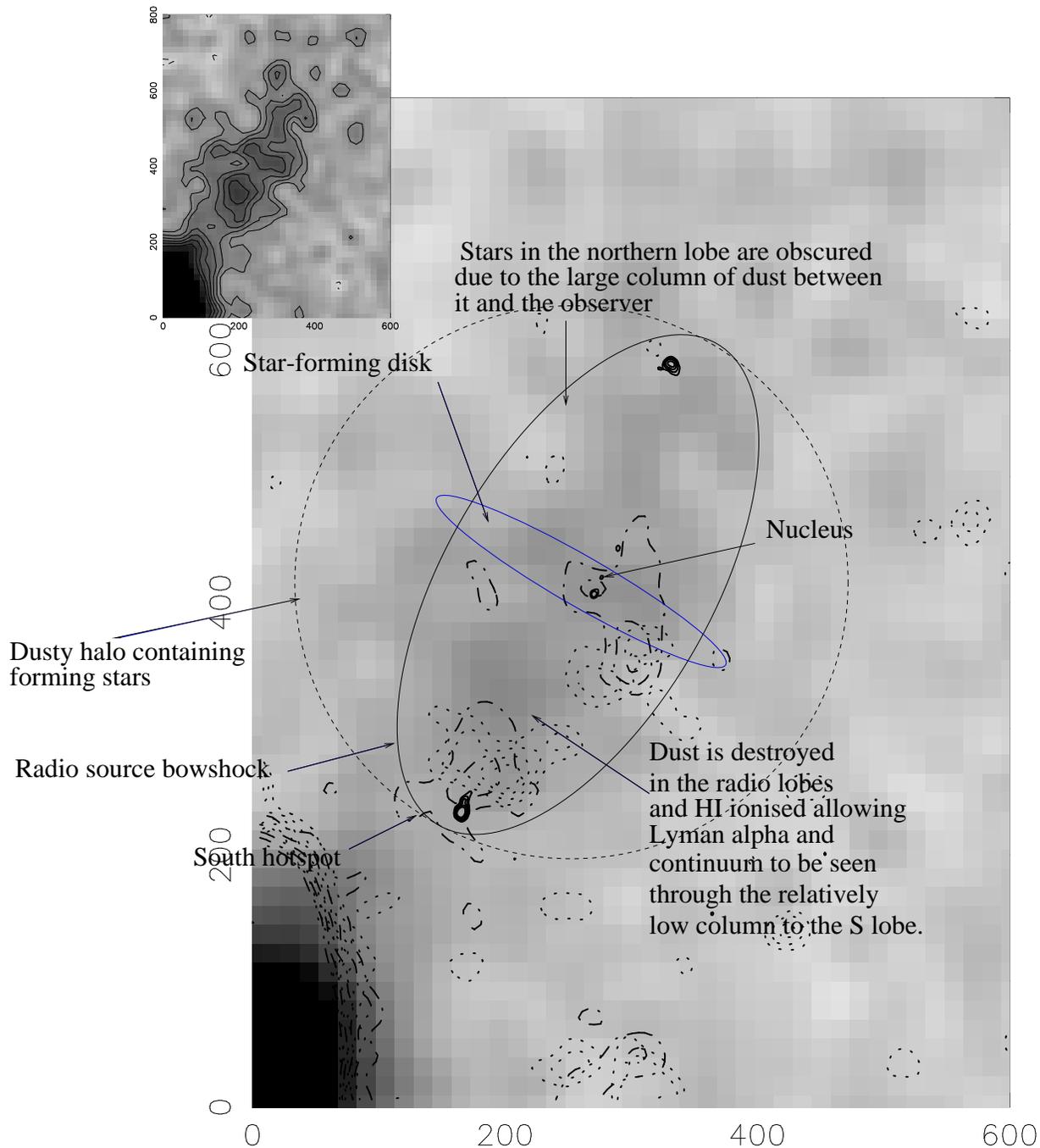}}
\end{picture}
\caption{Overlay of the F814W image (dot-dash contours), the MERLIN 
5GHz radio map (solid contours) and pseudo-Ly$\alpha$
image (dotted contours) over the smoothed
$K$-band image of 8C1435+635. Superposed is a cartoon depicting the model
discussed in the text 
(in which the south lobe is the closer to us). The image
is $6\times 8$-arcsec in size. {\em Inset:} the $K$-band image of 8C1435+635
smoothed with a $\sigma=0.15$-arcsec gaussian 
(contours and greyscale). Note the bar running across the middle of 
the source perpendicular to the radio axis.}  
\end{figure}

\subsection*{6C 0140+326}

One orbit in each of the F675W and F658N filters was obtained on this 
object. Both these filters contain the Ly$\alpha$ line, and the emission
in the F675W filter was found to consist entirely of line emission. In 
addition, NICMOS imaging of a nearby high-$z$ field galaxy
has also given us a high signal:noise NIC3 image of 6C0140.
The Ly$\alpha$ image obtained shows that the line emission is only from 
the western end of the radio source. 
A UV continuum image from the WHT (Fig.\ 2) shows 
marginally-detected UV continuum from most of the area of the radio source, 
and the $K$-band image (van Breugel et al.\ 1998) is bar-shaped and aligned 
along the radio axis. The object is being weakly lensed
by a nearby galaxy, and this is probably responsible for most of the 
elongation in the $K$-band, as well as the sinuous structure seen in the 
NIC3 image, although there is probably also 
some intrinsic radio--optical alignment.

\begin{figure}
\begin{picture}(400,250)
\put(-180,-480){\includegraphics{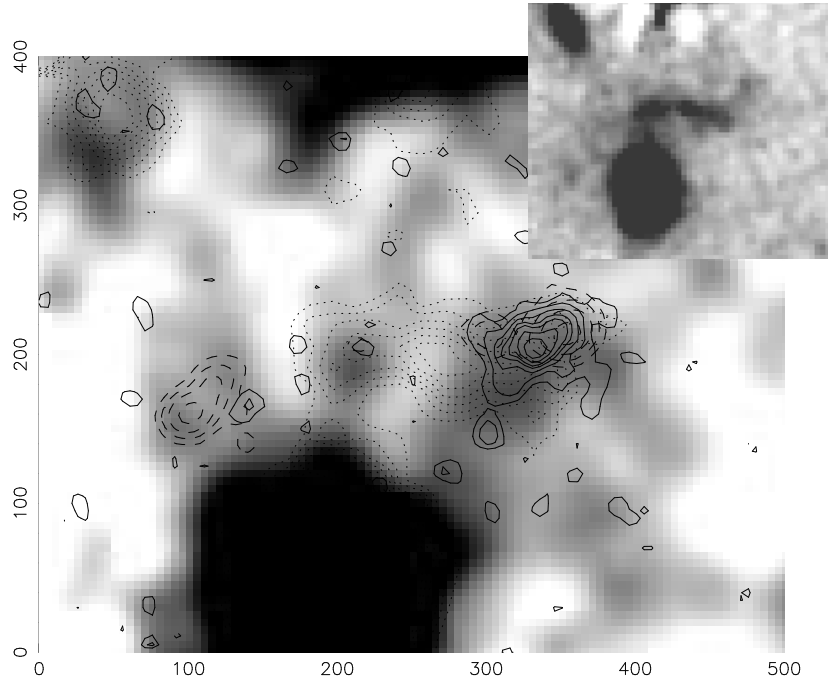}}
\end{picture}
\caption{Overlay of the 1.5 GHz MERLIN radio image (dashed contours), 
Ly$\alpha$ emission (solid contours) and
$K$-band emission (dotted contours) on a greyscale of an $I$-band WHT image of 
6C0140+326. Radio--optical registration has been obtained by lining up the 
putative radio central component with the peak in the UV/$K$-band emission.
The image is $5\times 4$-arcsec in size. {\em Inset:} the NIC3 F160W image
as a greyscale. In both images the bright star to the north has been 
subtracted.}
\end{figure}

\section*{A galaxy-scale starburst in 8C 1435+635?}
The submm detection of 8C 1435+635 (Ivison et al.\ 1998) implies the 
presence of $\sim 2\times 10^8 M_{\odot}$ of dust. Although this may be 
concentrated in the nucleus, as is the case for local ULIGS it may 
alternatively be distributed on the scale of the host galaxy ($\sim 40$ kpc), 
and be heated by a galaxy-wide starburst. If true,
this could mean that we are seeing a giant elliptical galaxy forming according
to a monolithic collapse scenario. 

Dust distributed throughout the galaxy would account for the 
relatively red UV-optical colour and the patchy nature of the 
UV continuum emission. Distributed
dust with H{\sc i} would also explain the peculiar distribution of the 
Ly$\alpha$ emission (Fig.\ 1) and the complicated velocity structure of the 
Ly$\alpha$ line (Lacy et al.\ 1994).

The UV flux and colour of 8C1435+635 is consistent with a starburst: the 
UV continuum flux corresponds to a star formation rate of 
$\approx 170 M_{\odot} {\rm yr}^{-1}$ {\em before correction for reddening}, 
if all the UV is from a starburst. The $F814W-K$ colour corresponds to 
a spectral index in $f_{\lambda}$ of $\beta \approx -1.1$. Using the models of
Meurer, Heckman \& Calzetti (1999), this corresponds to an extinction at 
160nm of 2.2 mag and an infrared:ultraviolet luminosity ratio 
of $\approx 10$. This is consistent with the far-infrared luminosity 
of this object as detected by SCUBA ($4.4\times 10^{12}L_{\odot}$), and 
implies a {\em corrected} star formation rate of $\sim 1200 M_{\odot}$
yr$^{-1}$. A good test of the distributed starburst 
theory would be to map the neutral gas directly using CO emission from the 
galaxy. This should be within the capability of the proposed large mm-array.

Of course, we must remember though that we are looking at a 
radio galaxy, and the radio source is likely to be influencing what
we are seeing. The aligned continuum in 8C1435+635 could represent,
for example, the result of dust scattering of a hidden quasar nucleus or 
inverse Compton emission from low energy electrons in the radio 
lobes (Spinrad et al.\ 1995). Nevertheless, the $K$-band
emission bar across the centre of the source would seem to lie 
outside any plausible scattering cone, and is therefore probably starlight.

Jet-induced star formation may also be contributing to the aligned
light, and the colour
gradient in the SE lobe could reflect an age gradient in the stellar 
population formed behind the expanding bowshock. 
A colour gradient in the aligned emission could also arise 
if the radio source is embedded in a dusty halo
of young stars, however. Shocks associated with the expanding radio source
may destroy dust grains within the radio lobes, resulting in bluer
emission from the end of the 
approaching lobe, as the column density of dust on the line 
of sight will be the lowest here. This model (Fig.\ 1) also provides a nice
explanation for the lack of UV emission from the northern lobe.

\acknowledgments
I thank my collaborators on the HST project, Hy Spinrad, Arjun Dey 
and Steve Rawlings. 
I am grateful to Wil van Breugel and Carlos de Breuck for supplying 
their $K$-band images of 6C 0140+326 and 8C 1435+635, and to Andy 
Bunker for putting up this poster in my absence.


\begin{references}
\reference Ivison R.J., et al., 1998, ApJ, 494, 211  
\reference Lacy M., et al., MNRAS, 271, 504 
\reference Meurer G.R., Heckman T.M., Calzetti D., 1999, ApJ, in press
\reference Rawlings S., Lacy M., Blundell K.M., Eales S.A., Bunker A.J., Garrington S.T., 1996, Nat, 383, 502
\reference Spinrad H., Dey A., Graham J., 1995, ApJ, 438, 51
\reference van Breugel W.J.M., Stanford S.A., Spinrad H., Stern D., Graham J.R., 1998, ApJ, 502, 614

\end{references}
\end{document}